\documentclass[aps,prb,citeautoscript,twocolumn,amsmath,amssymb,reprint,superscriptaddress,floatfix
]{revtex4-1}
\usepackage{graphicx}
\usepackage{epstopdf} 
\usepackage{siunitx}
\usepackage{amsmath} 
\usepackage{romannum}
\usepackage{blindtext}
\usepackage{color}

\newcommand{\etal}{\textit{et al}.\ }
\DeclareSIUnit\dBm{dBm}
\DeclareSIUnit\dB{dB}
\DeclareSIUnit\mbar{mbar}
\newcommand{\RNum}[1]{\uppercase\expandafter{\romannumeral #1\relax}}

\usepackage{filecontents}

\begin{document}

\title{Quantised conductance of one-dimensional strongly-correlated electrons in an oxide heterostructure}

\author{H. Hou}
 \affiliation{ 
Cavendish Laboratory, University of Cambridge, CB3 0HE, UK.}
\author{Y. Kozuka}
\affiliation{
Department of Applied Physics and Quantum-Phase Electronics Center (QPEC), The University of Tokyo,
Tokyo 113-8656, Japan
}

\affiliation{
Research Center for Magnetic and Spintronic Materials, National Institute for Materials Science (NIMS), 1-2-1 Sengen, Tsukuba 305-0047, Japan
}
\affiliation{
JST, PRESTO, Kawaguchi, Saitama 332-0012, Japan
}
\author{Jun-Wei Liao}
\author{L. W. Smith}
\author{D. Kos}
\author{J. P. Griffiths}%
\affiliation{ 
Cavendish Laboratory, University of Cambridge, CB3 0HE, UK.}
\author{J. Falson}%
\affiliation{
Max Planck Institute for Solid State Research, D-70569 Stuttgart, Germany
}
\author{A. Tsukazaki}%
\affiliation{
Institute for Materials Research, Tohoku University, Sendai 980-8577, Japan
}
\author{M. Kawasaki}%

\affiliation{
Department of Applied Physics and Quantum-Phase Electronics Center (QPEC), The University of Tokyo, Tokyo 113-8656, Japan
}
\author{C. J. B. Ford}%
\email{cjbf@cam.ac.uk}
\affiliation{
Cavendish Laboratory, University of Cambridge, CB3 0HE, UK.
}
\begin{abstract}
Oxide heterostructures are versatile platforms with which to research and create novel functional nanostructures. We successfully develop one-dimensional (1D) quantum-wire devices using quantum point contacts on MgZnO/ZnO heterostructures and observe ballistic electron transport with conductance quantised in units of $2e^2/h$. Using DC-bias and in-plane field measurements, we find that the $g$-factor is enhanced to around 6.8, more than three times the value in bulk ZnO. We show that the effective mass $m^*$ increases as the electron density decreases, resulting from the strong electron-electron interactions. In this strongly interacting 1D system we study features matching the `0.7' conductance anomalies up to the fifth subband.
This paper demonstrates that high-mobility oxide heterostructures such as this can provide good alternatives to conventional III-V semiconductors in spintronics and quantum computing as they do not have their unavoidable dephasing from nuclear spins. This paves a way for the development of qubits benefiting from the low defects of an undoped heterostructure together with the long spin lifetimes achievable in silicon.

\end{abstract}

\maketitle
\preprint{}
\section{Introduction}
Physical phenomena in transition-metal oxides and their complex compounds have stimulated intense interest in research covering metallic, semiconducting, and insulating properties. At the heterointerface of two oxide layers, the symmetry breaking leads to novel properties including quantum confinement of electrons, strong correlations, superconductivity, and ferromagnetism. \cite{zubko2011interface}
In a MgZnO/ZnO heterostructure, the polarisation mismatch between MgZnO and ZnO originating from spontaneous and piezoelectric contributions induces a two-dimensional electron gas (2DEG) at the interface. \cite{tsukazaki2007quantum}
By engineering the strain via Mg composition and the MgZnO thickness, the 2DEG density can reach down to $10^{11}$\,cm$^{-2}$ with mobility over $10^6$\,cm$^2$V$^{-1}$s$^{-1}$ \cite{falson2016mgzno}.
Both integer and fractional quantum-Hall effects have been investigated\cite{tsukazaki2007quantum,tsukazaki2010observation,falson2015even}. Furthermore, the weak spin-orbit interaction and the low concentration of nuclear spins in ZnO lead to a long spin-coherence time\cite{kozuka2013rashba}. These unique properties in the MgZnO/ZnO heterostructure create an excellent platform for investigating quantum physics beyond the more conventional \RNum{3}-\RNum{5} alternatives\cite{kozuka2014challenges,0034-4885-81-5-056501}. More recently, Andreev reflection has been demonstrated at the interface between a MgZnO/ZnO heterostructure and MoGe superconductor, and this could be a good system for investigating non-abelian quasiparticles, such as Majorana fermions.\cite{kozuka2018andreevref}

Most low-dimensional devices use gates to define nanostructures such as one-dimensional (1D) quantum point contacts (QPCs)/quantum wires or quantum dots. However, gating oxide heterostructures is a challenge. In a MgZnO/ZnO heterostructure, we have found that using standard insulators such as Al$_{2}$O$_{3}$ causes a reduction in mobility and strong parallel conduction. The latter may be because the hard Al$_{2}$O$_{3}$ deposited on the thin stressed MgZnO layer contributes to a mismatch of the spontaneous and piezoelectric polarisations, and induces another 2DEG at the Al$_{2}$O$_{3}$/MgZnO interface. So far it has been necessary to use one-off techniques to create nanostructures, such as conducting AFM lithography on LaAlO$_{\rm 3}$/SrTiO$_{\rm 3}$, which showed quantised conductance in units of $e^2/h$ in a strong magnetic field \cite{cen2009oxide,tomczyk2016micrometer,annadi2016quantized}. In our work, we have successfully prevented parallel conduction by replacing hard Al$_{2}$O$_{3}$ insulator with soft parylene C.

The zero-field quantisation of conductance in integer multiples of $2e^2/h$ is a signature of ballistic charge transport in 1D systems. The lateral electrostatic confinement creates a series of 1D subbands, in which spin-up ($\uparrow$) and spin-down ($\downarrow$) subbands each contribute $e^2/h$. 
This is already observed in many materials, including GaAs/AlGaAs \cite{wharam1988one, vanwees1988}, InGaAs/InAlAs heterostructures\cite{bever1994quantized}, strained epitaxial germanium\cite{gul2017quantum} and carbon-based materials \cite{frank1998carbon,tombros2011quantized}. An anomalous feature at conductance $G=0.7\times 2e^2/h$ was first investigated by Thomas \etal and attributed to a possible spontaneous spin polarisation \cite{thomas1996possible, rokhinson2006spontaneous}. Its origin has since been much debated \cite{micolich2011review}, and other explanations proposed including quasi-bound-state formation and the Kondo effect \cite{cronenwett2002kondo, meir2002kondo}. Recently, Bauer \etal used a smeared van Hove singularity to explain it and emphasised the important role that electron-electron interactions play in the 0.7 anomaly\cite{bauer2013microscopic}.

Here we report ballistic transport in a high-quality MgZnO/ZnO heterostructure and show well-defined conductance quantisation. Using DC-bias spectroscopy \cite{patel1991evolution} and in-plane magnetic-field measurements, we find a $g$-factor in the 1D wire that is enhanced by a factor of $\sim 3$ compared to the bulk and is fairly constant at low 1D subband index. Additionally, we show that the effective mass $m^*$ increases as the density decreases in our 1D system, as occurs for 2DEGs in similar heterostructures\cite{kasahara2012correlation,falson2015electron}. 

\begin{figure}[t]
\includegraphics[width=\columnwidth]{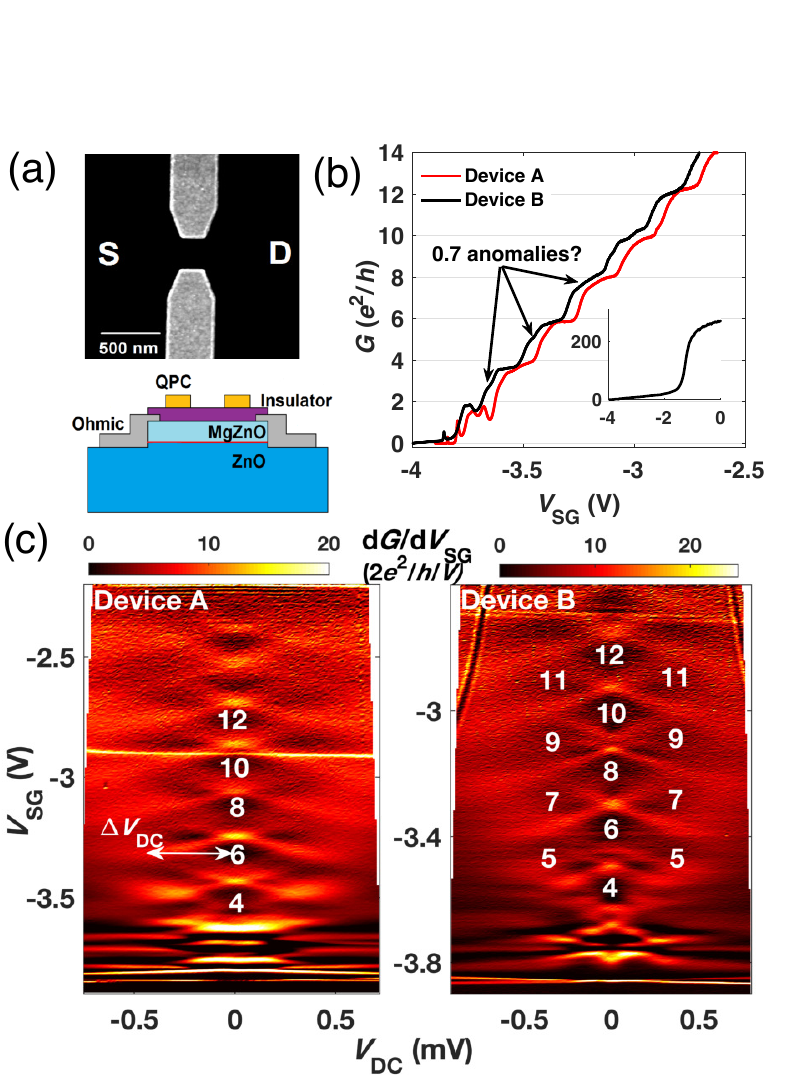}
\caption{(a) Scanning electron micrograph of a QPC (source S and drain D), and schematic diagram of the device cross-section across the channel. The 2DEG forms at the MgZnO/ZnO interface (red line). Ti/Au gates are patterned using electron-beam lithography and deposited on parylene-$C$ insulator.
(b) Conductance $G$ as a function of split-gate voltage $V_{\rm SG}$ (devices A and B), showing plateaus at multiples of $2e^2/h$. Inset: $G$ over the full $V_{\rm SG}$ range. The drop corresponds to depletion of electrons beneath the gates leaving a quasi-1D wire in the gap.
(c) Transconductance ${\rm d}G/{\rm d}V_{\rm SG}$ \textit{vs} bias $V_{\rm DC}$ and $V_{\rm SG}$ (devices A and B). Dark (bright) regions correspond to plateaus (risers) in conductance. The energy difference between the third and fourth 1D subbands is $e\Delta V_{\rm DC}$, which is measured from the crossing point of adjacent subbands as indicated in the plot. The numbers indicate heights of conductance plateaus (units $e^2/h$).}
\label{setup}
\end{figure}
The strongly correlated electron system in MgZnO/ZnO heterostructures offers a novel platform to investigate interaction effects, particularly the 0.7 structure.
These strong correlations arise from the high $m^*$ ($=0.3\,m_{\rm e}$ in bulk ZnO, where $m_{\rm e}$ is the bare electron mass) and small dielectric constant $\varepsilon=8.3$ compared to GaAs. The ratio $r_{\rm s} = E_{\rm C}/E_{\rm F}$ can reach 10 in low-density wafers ($E_{\rm C}$ is the average Coulomb energy per electron and $E_{\rm F}$ is the Fermi energy). These strong interactions may be the source of `$N.7$' structures in 1D subbands with higher index $N$ apparent in our data, which behave like the 0.7 structure. Any such $N.7$ structures  \cite{thomas1998interactioneffects,Chris07,koop2007,danneau20080} are very weak in GaAs electron or hole systems. We confirm the importance of electron-electron interactions in the 0.7 anomaly with an independent measurement of the strength of the electron-electron interaction from the electron effective mass.\cite{bauer2013microscopic}


\section{Experimental details}

The MgZnO/ZnO heterostructures are grown by molecular-beam epitaxy\cite{falson2016mgzno}. Devices A and B (C) have a 2DEG \SI{92}{nm} (\SI{500}{nm}) below the surface, density \SI{3.2e11}{cm^{-2}} (\SI{1.2e12}{cm^{-2}}) and mobility \SI{3.5e5}{cm^2V^{-1}s^{-1}} (\SI{6e4}{cm^2V^{-1}s^{-1}}), corresponding to an electron mean free path of $l_{\rm e}=\SI{3.2}{\mu m}$ (\SI{1.1}{\mu m}). A scanning electron micrograph and schematic cross-section of a device are shown in Fig.~\ref{setup}(a). A Hall-bar mesa is patterned by Ar ion milling and Ti/Au is thermally evaporated to create Ohmic contacts without annealing. Ti/Au split gates of length $L=\SI{200}{nm}$ and width $ W=\SI{300}{nm}$ are deposited on a \SI{30}{nm}-thick parylene-$C$ insulator layer, forming a quasi-1D wire in the 2DEG. We measure conductance through the QPC using a four-terminal lock-in technique with excitation voltage \SI{10}{\mu V} at \SI{77}{Hz}, at a temperature of $\sim$\SI{50}{mK}. 
For perpendicular magnetic-field measurements we measure the diagonal resistance to obtain the number of transmitted Landau-level edge states\cite{buttiker1988absence}.

\section{Results and Discussion}

\subsection{Quantised 1D conductance}
Figure~\ref{setup}(b) shows the 1D conductance \textit{vs} split-gate voltage ($V_{\rm SG}$). At $V_{\rm SG}=\SI{-1.5}{V}$, electrons below the gate are depleted and a quasi-1D wire is defined in the gap [inset, Fig. \ref{setup}(b)]. This definition voltage matches the expected value calculated using a parallel-plate capacitor model with this 2DEG density, dielectric constant and thickness. 
1D conductance plateaus appear as $V_{\rm SG}$ is made more negative, quantised in units of $\Delta G\approx 2e^2/h$. They are visible up to \SI{14}{e^2/h} (devices A and B). For $G<2e^2/h$, Coulomb-blockade (CB) peaks appear, probably because dots form owing to the reduced electron screening increasing disorder, so we will discuss results from the second plateau and above.

Figure~\ref{setup}(c) shows the transconductance ${\rm d}G/{\rm d}V_{\rm SG}$ \textit{vs} source-drain bias $V_{\rm DC}$ (devices A and B). Dark (light) regions correspond to plateaus (transitions between plateaus). The splitting of the source and drain chemical potentials causes the risers to split until quantised plateaus appear between them at odd-integer values $G=Me^2/h$ ($M=3,5,7\dots$), giving diamond-shaped features in Fig.~\ref{setup}(c)~\cite{patel1991evolution}. The 1D subband energy spacing $\Delta E$ is given by the size of the diamond $\Delta V_{\rm DC}$ times $e$, and is around \SI{0.4}{meV} here. Surprisingly, this remains reasonably constant as the 1D subband index decreases.
Because of the high $m^*$ and hence relatively small subband spacing, the plateaus are strongly temperature dependent, disappearing for $T\gtrsim \SI{1}{K}$ (Fig.~S1 in Supplemental Material (SM)). 
\subsection{Magnetic-field dependence \& Enhancement of $g$-factor}

\begin{figure}[ht]
\includegraphics[width=\columnwidth]{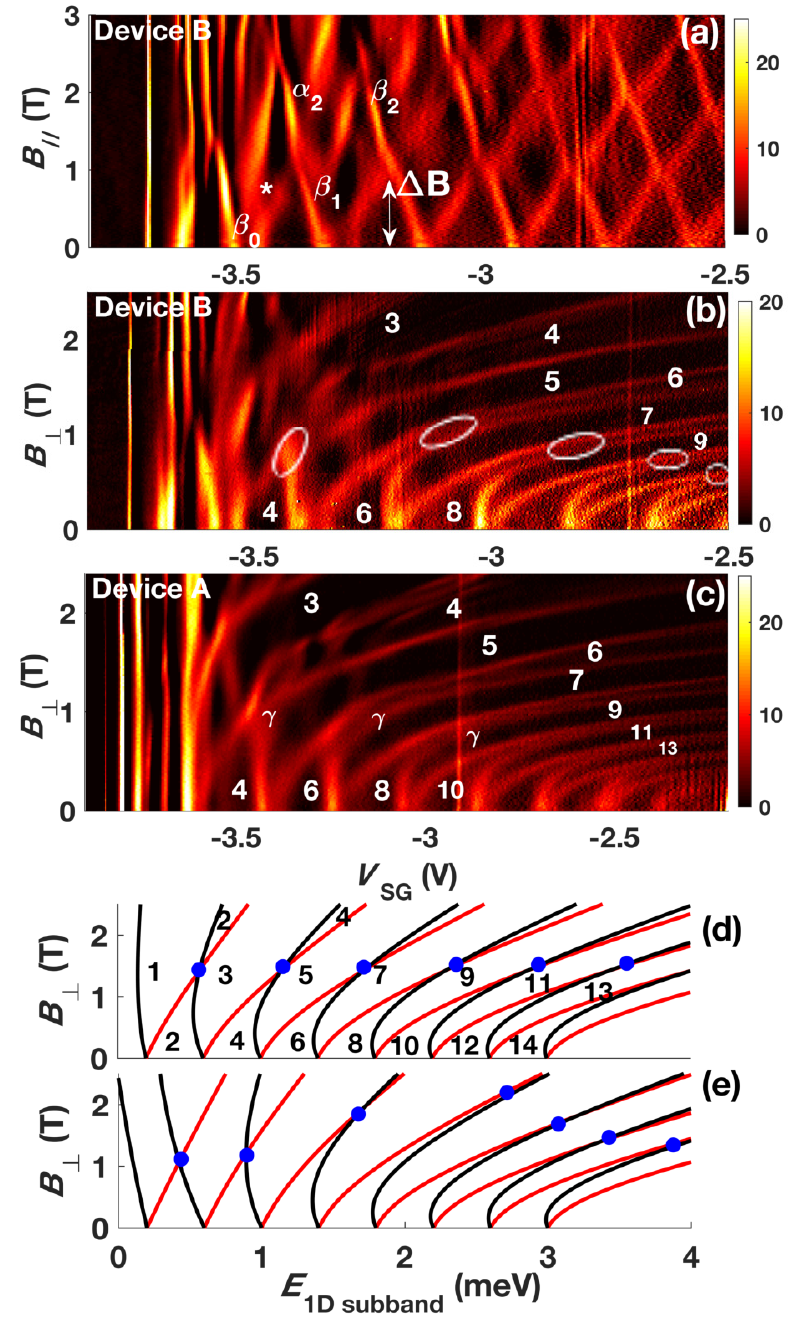}

\caption{The differential conductance ${\rm d}G/{\rm d}V_{\rm SG}$ (units $2e^2/h/V$) \textit{vs} $V_{\rm SG}$ at different (a) $B_{\|}$, and (b) and (c) $B_{\perp}$ for devices as labelled. $\Delta B$ indicates the required $B_{\|}$ for the crossing of subbands $2_\uparrow$ and $3_\downarrow$. Ellipses in (b) indicate positions of subband crossings. The calculated electron energy spectrum \textit{vs} $B_{\perp}$ with (d) constant and (e) increasing $m^*$. The numbers indicate the heights of conductance plateaus (units $e^2/h$), and blue dots mark positions of crossings.}
\label{outplane}
\end{figure}

In an in-plane field, the quantised plateaus at even multiples of $e^2/h$ split due to the Zeeman energy $E_{\rm Z}$ (Fig.~\ref{outplane}(a)). At $B_{\|}=\SI{0.5}{T}$, plateaus occur at $G=ne^2/h$, ($n=3,4,5\ldots$), when a 1D subband of the lower-energy spin-polarisation direction becomes fully transmitted before the subband with opposite spin. As $B_{\|}$ increases further to \SI{1}{T}, the spin-split 1D subbands cross, leaving plateaus only at $G=Me^2/h$ ($M=3,5,7\ldots$). From $E_{\rm Z}$ and the subband spacing, we estimate the $g$-factor as\cite{patel1991evolution}
\begin{equation}
\vert g^*\vert=\frac{1}{\mu_{\rm B}}\frac{\delta E}{\delta V_{\rm SG}}\frac{\delta V_{\rm SG}}{\delta B}=\frac{e}{\mu_{\rm B}}\frac{\Delta V_{\rm DC}}{\Delta B},
\label{gfactor}
\end{equation} 
with $e$ the electronic charge and $\mu_{\rm B}$ the Bohr magneton.
We estimate $g^{*} \approx 6.8$ (6.4) for the second (fifth) subband, enhanced above the bulk value $g^*\approx 2$ for heterostructure and bulk ZnO\cite{kozuka2013rashba}. An enhancement of $|g^{*}|$ by a similar factor occurs for electrons in GaAs QPCs, with values from 0.75 to 1.5\cite{patelmagneticfield, thomas1996possible, cronenwett07}, compared to 0.44 in the bulk. For electrons in GaAs QPCs, $|g^*|$ decreases fairly rapidly with subband index\cite{thomas1998interactioneffects}, whereas in our data $g^{*}$ is almost constant, as the Zeeman splittings of subbands are almost identical [Fig.~\ref{outplane}(a)]; the gradient of subband edges and the field at which subbands cross are very similar for subbands 2 to 6.

For a field applied perpendicular to the 2DEG ($B_{\perp}$), the electron energy contains Zeeman ($E_{\rm Z}=g^*\mu_{\rm B}B$) and cyclotron-energy ($E_{\rm c}=\hbar\omega_{\rm c}$) terms:
\begin{equation}
E_{\rm N}(B)=\left(N+\frac{1}{2}\right)\hbar\sqrt{\omega_{\rm 0}^2+\omega_{\rm c}^2}\pm\frac{1}{2}g^*\mu_{\rm B}B,
\label{subspace}
\end{equation}
($N$ is subband index, $\hbar\omega_{\rm 0}$ subband spacing assuming a parabolic potential, and $\omega_{\rm c}=eB_{\perp}/m^*$ the cyclotron frequency).\cite{berggren1986} 

In the low-field regime ($\hbar\omega_{\rm 0} \gg \hbar\omega_{\rm c}$), spin-degenerate plateaus at even multiples of $G=e^2/h$ are split by $E_{\rm Z}$, leading to additional plateaus at $G=ne^2/h$ ($n=3,4,5\ldots$). These merge to odd multiples of $G=e^2/h$ with increasing $B_{\perp}$ as adjacent spin-split subbands cross. However, for high $B_{\perp}$, Landau-level formation leads to the creation of hybrid magneto-electric subbands in the constriction \cite{berggren1986, vanwees1988, wharam1988one, srinivasan2013}, for which plateaus again occur at both even and odd integers. The onset of this regime can be quantified by the ratio $\kappa=\hbar\omega_{\rm c}/E_{\rm Z}= \hbar e/(g^*m^*\mu_{\rm B})$. We estimate $m^*=(0.36\pm0.02) \,m_{\rm e}$ from temperature-dependent Shubnikov--de-Haas measurements and the Dingle formula \cite{coleridge1991dingle} (Fig.~S2 in SM). This gives $\kappa<1$ for $g^*=6.8$. The smaller $m^* = 0.067\,m_{\rm e}$ and $g$-factor ($\approx 1$) in GaAs devices leads to $\kappa \gg 1$, so odd-integer plateaus are not observed.

Figures~\ref{outplane}(b) and (c) show the evolution of subbands with $B_{\perp}$. Bright regions correspond to risers in conductance between plateaus, where subband edges cross the Fermi energy. Plateau heights are labelled. We model the subband energy \textit{vs} $B$ using Eq.~\ref{subspace}. Results for constant $m^* = 0.4\,m_{\rm e}$, $g^*$ and $\hbar \omega_0$ are plotted in Fig.~\ref{outplane}(d), black and red lines representing spin-down ($\downarrow$) and spin-up ($\uparrow$) subbands, respectively. The pattern of plateaus matches the experiment well. However, the value of $B_{\perp}$ at which subbands cross is independent of subband index, which does not match the measurements. In Fig. \ref{outplane}(b), the crossing between $N=2_\uparrow$ and $N=3_\downarrow$ subbands occurs around $B_{\perp}=\SI{0.8}{T}$, and between $N=3_\uparrow$, $4_\downarrow$ around $B_{\perp}=\SI{1.1}{T}$. For subbands $N>4$, the required $B_{\perp}$ decreases. 

Given that $m^*$ in ZnO increases with decreasing density\cite{kasahara2012correlation,falson2015electron}, we repeat the calculation with $m^*$ increasing from 0.4 to $1\,m_{\rm e}$ as the subband index decreases, (Fig.~\ref{outplane}(e)). This reproduces the trend from experimental data that the value of $B_{\perp}$ at which spin-split subbands cross initially increases, then decreases for higher subbands (although the model tends to overestimate $B_{\perp}$, or perhaps experimental crossing points are underestimated due to energy blurring). We vary $m^*$ instead of $g^*$ since our data suggests that $g^*$ is reasonably independent of subband index, unlike in GaAs. In addition, DC-bias spectroscopy (Fig.~\ref{setup}(c)) indicates a reasonably constant subband spacing over this range. Experimentally, the precise points at which bands cross at low index cannot be determined, and some look more like anticrossings (labelled $\gamma$, Fig.~\ref{outplane}(c)). This may be because of a strong electron-electron interaction, and could be modelled by introducing an exchange interaction term in Eq.~\ref{subspace}\cite{daneshvar1997magnetization,daneshvar1998multiple}.

\subsection{Effective mass measurements}
\begin{figure}[ht]
\includegraphics[width=\columnwidth]{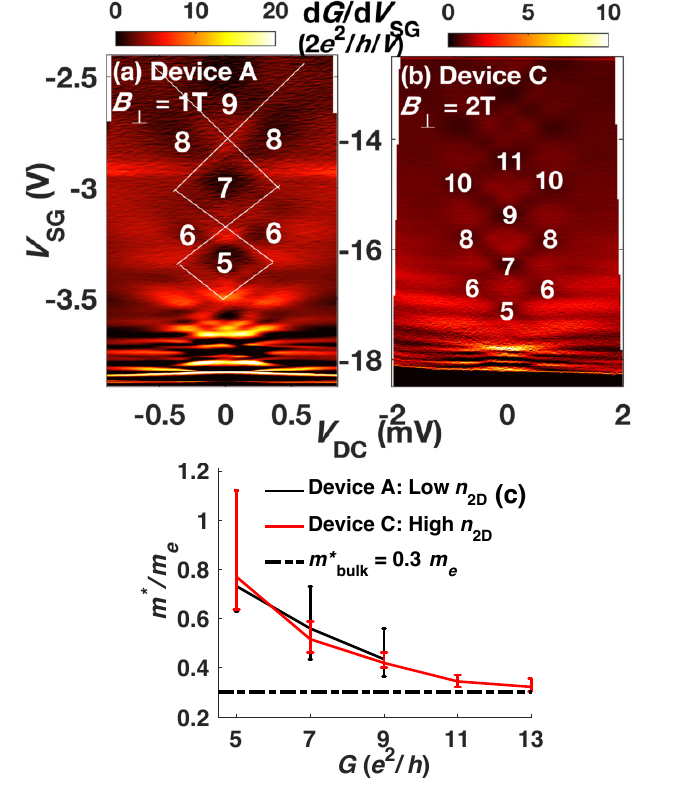}
\caption{Transconductance $dG/dV_{\rm SG}$ as a function of $V_{\rm SG}$ and $V_{\rm DC}$ for (a) device A at $B_{\perp}=\SI{1}{T}$, and (b) device C at $B_{\perp} = \SI{2}{T}$. In the device C, the high electron density and thick MgZnO (\SI{500}{nm}) require strong negative voltages to define the 1D wire. The plateau conductances are indicated (units $e^2/h$). (c) $m^*$ measured \textit{vs} $G$ in both devices, compared with the bulk value for ZnO.}
\label{effective}
\end{figure}
To investigate $m^*$ further we measure the DC-bias dependence at $B_{\perp}=\SI{1}{T}$ (device A) (Fig.~\ref{effective}(a)). Adjacent spin-split subbands cross near this value of $B_{\perp}$, indicated by only odd plateaus being present
for $V_{\rm DC}=0$. Since the subband spacing is roughly constant, each pair of subbands $2_{\uparrow}$/$3_{\downarrow}$, $3_{\uparrow}$/$4_{\downarrow}$ etc is degenerate since spin $\uparrow$/$\downarrow$ subbands are shifted by $+$/$-$ $\frac{1}{2}g\mu_{\rm B} B$, respectively, cancelling out $E_{\rm Z}$. The energy difference between these pairs of subbands is
$
\Delta E=\hbar\sqrt{\omega_{\rm 0}^2+\omega_{\rm c}^2}.
$
In contrast to the $B=0$ case, at $B_{\perp}=\SI{1}{T}$ the spacing between subband pairs decreases at lower subband index $N$ (Fig.~\ref{effective}(a)). This could be explained by increasing $m^*$ at lower density, leading to a smaller $E_{\rm c}=\hbar eB/m^*$ for lower $N$.
Fig.~\ref{effective}(b) shows the measurement repeated for device C with different $n_{\rm 2D}$ and device dimensions ($L=\SI{300}{nm}$ and $W=\SI{800}{nm}$), in which more plateaus are evident. The same trend of increasing spacing with subband index occurs.

Figure~\ref{effective}(c) shows $m^*$ \textit{vs} plateau height (units $e^2/h$), estimated using $\Delta E$. At high conductance $(\sim 13 e^2/h)$, $m^*=(0.31\pm 0.03)\,m_{\rm e}$, close to the bulk effective mass found above for this wafer. When $G$ decreases to $5e^2/h$, $m^*$ increases to $(0.96\pm 0.2)\,m_{\rm e}$, which is comparable to that of a 2D system at a lower density of \SI{1e11}{cm^{-2}} ($0.8\pm 0.2)\,m_{\rm e}$. \cite{falson2015electron} The large error bar is due to the blurred nature of this subband crossing. However, the trend of increasing $m^*$ as $G$ decreases is clear. 

In previous studies of MgZnO/ZnO heterostructures, $m^*$ measured from temperature-dependent Shubnikov--de-Haas oscillations increases as the 2DEG density decreases, while $m^*$ from cyclotron resonance is roughly constant\cite{kasahara2012correlation,falson2015electron}. This indicates that the increase in transport effective mass arises from electron-electron interactions, which are more significant at lower density. A similar effect is observed in other 2DEG systems,\cite{GaAseffectivemass,pudalov2002low} but much weaker (a factor of $\sim 1.4$ rather than 3 as in these ZnO heterostructures). In 1D GaAs wires, an increase in $m^*$ by at most 30\%
was reported when the 1D density decreased from \SI{2.6e8}{m^{-1}} to \SI{1e8}{m^{-1}}.\cite{ploner1998energy} However, the ratio of electron-electron interaction energy to kinetic energy in GaAs is relatively low, and non-parabolicity, disorder and electron-phonon interactions may also contribute significantly to this increase.\cite{degani1988interaction}
\begin{figure}[ht]
\includegraphics[width=\columnwidth]{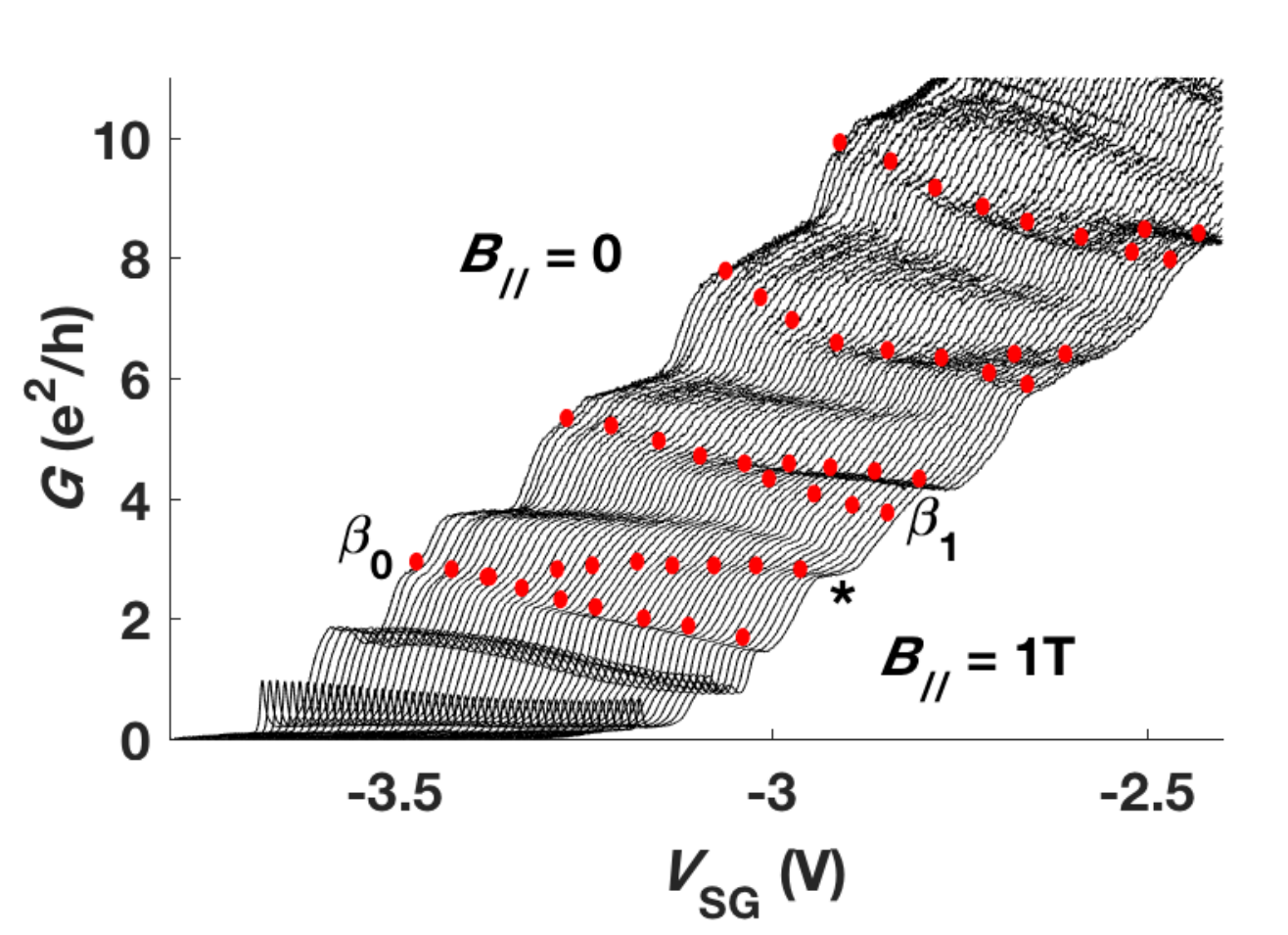}
\caption{Conductance $G$ \textit{vs} $V_{\rm SG}$ at different $B_{\|}$ from 0 to \SI{1}{T} in steps of \SI{0.02}{T}. Each trace is shifted to the right by \SI{0.01}{V} relative to the previous trace for clarity. Red dots illustrate how the shoulder features resembling the 0.7 anomaly evolve with $B_{\|}$ (same measurement data and labels as in Fig.~\ref{outplane}(a)). An estimated series resistance of around \SI{110}{\Omega} has been subtracted.}
\label{DL}
\end{figure}
\subsection{0.7 anomaly}
We now return to our discussion on the 0.7 anomaly. Fig.~\ref{setup}(b) shows several shoulder-like features below the main plateaus. We test whether they behave similarly to the 0.7 anomaly or to CB-like resonances from impurities.

i) Resonant peaks from CB should split with $V_{\rm DC}$ at a rate determined by the size of the dot-like impurity and coupling to the gates, but DC-bias spectroscopy for our devices shows an orderly splitting above the second plateau typical of clean 1D devices \cite{kristensen2000dcbias, chen2010direct}, [Fig.~\ref{setup}(c)].

ii) In Figs.~\ref{outplane}(a-c), even low $B$ lifts spin-degeneracy. The 
edges of spin-split subbands do not meet at $B=0$, showing clear gaps [for example, $\beta_{0}$ in Fig.~\ref{outplane}(a)] at the plateau, as seen for the 0.7 anomaly in GaAs.\cite{thomas1996possible}
Fig.~\ref{outplane}(a) also shows gaps at higher-order crossings (labelled $\alpha_{2}$, $\beta_{1}$, $\beta_{2}$ matching labels in Ref.~\citenum{graham2003analog}), which is an important sign of the 0.7 analogue\cite{graham2003analog}. The gaps can be explained as an effect of interactions.\cite{berggren2005analog}

iii) The conductance sweeps in Fig.~\ref{outplane}(a) are re-plotted (Fig.~\ref{DL}) as lines up to $B_{\|}=$\SI{1}{T}. The $N.7$ plateaus (just below $G=2(N+1)e^2/h$) appear to evolve smoothly to spin-polarised plateaus at $G=2(N+1/2)e^2/h$, then they split to form an extra plateau [indicated by * in Fig.~\ref{outplane}(a) and Fig.~\ref{DL}]. This split before the crossing was observed in GaAs\cite{graham2003analog}, but was much weaker. How interactions contribute needs further theoretical investigation. In Fig.~\ref{DL}, the $N.7$ structures strengthen and occur lower on the riser for lower subbands, consistent with more significant interactions\cite{bauer2013microscopic, smith2015} due to the lower density.

iv) Plateaus and $N.7$ structures stay reasonably constant as the wire is shifted laterally by asymmetric bias on the QPC gates (Fig.~S3 in SM).
The wire position should not significantly affect either quantisation or interaction effects inherent in thef 1D system such as the 0.7 structure.

While tests i) to iv) described above are not fully comprehensive, they give a strong indication that these structures belong to the 0.7 family, and the great similarities with the complex magnetic-field behaviour in GaAs are striking. An additional test is the temperature dependence. Because of the small 1D subband spacing in our samples, any structure is rapidly smeared by temperature, disappearing by $T>\SI{1}{K}$ (Fig.~S1 in SM).
$N.7$ structures are more visible in QPCs in MgZnO/ZnO compared to GaAs heterostructures because of the large effective mass and small dielectric constant leading to strong electron-electron interactions. From the cyclotron energy, we estimated the strength of electron-electron interactions using the electron effective mass. The increasing interaction energy relative to the kinetic energy explains why $N.7$ structures become better defined as the 1D subband index decreases, which is consistent with the model. \cite{bauer2013microscopic}
MgZnO/ZnO heterostructures show dilute ferromagnetic properties\cite{maryenko2017znoAHE} (see SM) that may help to enhance the local spin susceptibility in the channel at $B=0$\cite{bauer2013microscopic}.
\section{Conclusion}
To conclude, we have shown ballistic electron transport with conductance quantisation in 1D quantum wires defined on a MgZnO/ZnO heterostructure. We also find an increasing effective mass at lower density, consistent with measurements on 2D ZnO systems. Because of the large $g^*$ and $m^*$, a perpendicular field drives the system into a regime where the Zeeman energy is greater than the cyclotron energy, leading to only odd plateaus in the conductance. 
At zero field we see evidence of 0.7-like anomalies up to the fifth 1D subband. Such structures are rarely observed in GaAs, a key reason for which is probably the significantly higher interaction strength in ZnO. The ballistic transport and the importance of interactions and spin, together with a long spin-coherence time owing to the low concentration of nuclear spins in ZnO, could make high-quality MgZnO/ZnO heterostructures an interesting alternative to \RNum{3}-\RNum{5} semiconductors as platforms for quantum information and spintronic technologies.

\begin{acknowledgments}
We thank Stuart Holmes for helpful discussions. H.~Hou acknowledges the Chinese Scholarship Council and Cambridge Trust for financial support. This work was partly supported by JST, PRESTO Grant Number JPMJPR1763 and JST, CREST Grant Number JPMJCR16F1, Japan.
\end{acknowledgments}

\section{Supplement material}

\subsection{Temperature-dependence of the conductance}
\begin{figure}[h]
\includegraphics[width=0.7\columnwidth]{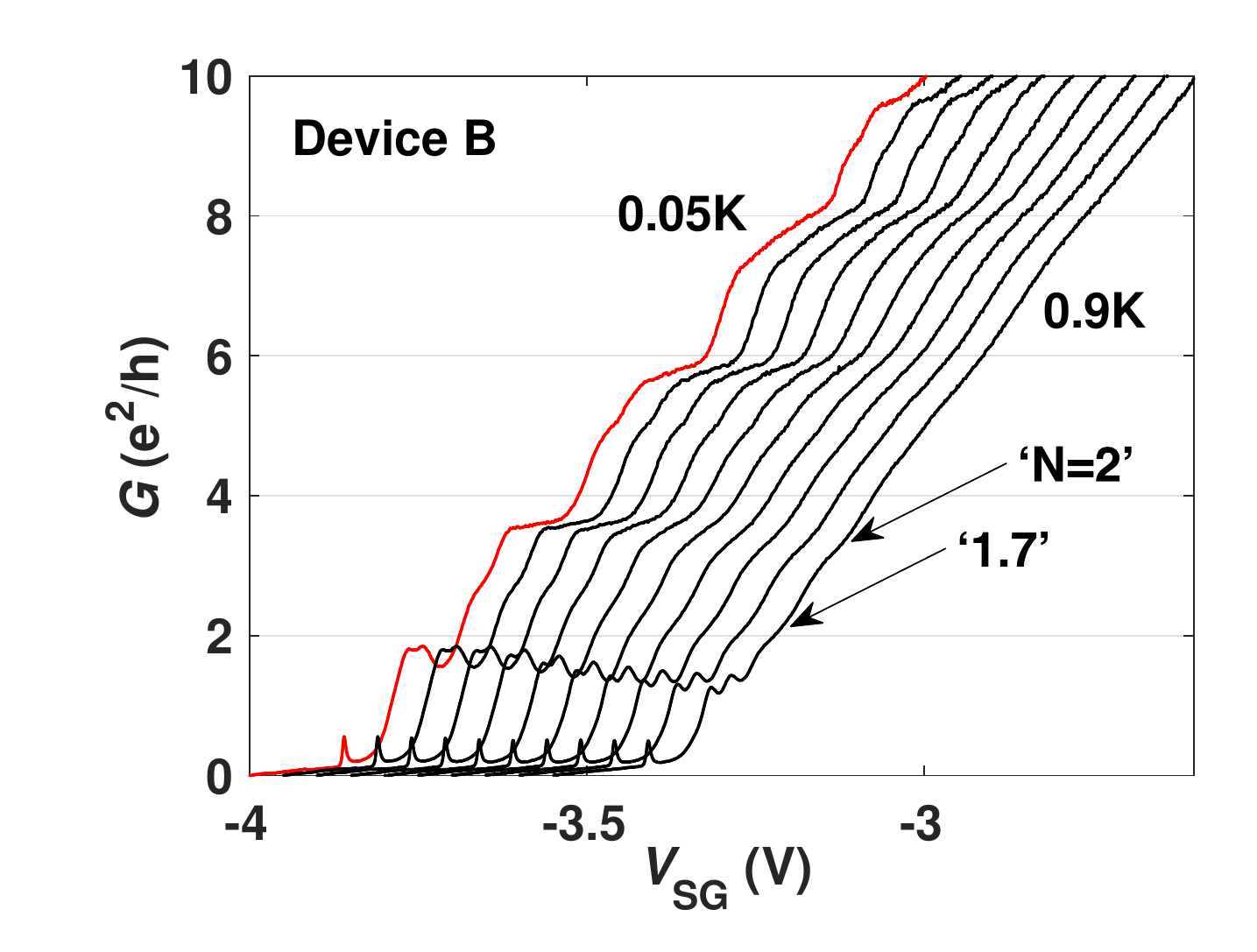}
\caption{Conductance $G$ as a function of split-gate voltage $V_{\rm SG}$ for device B, as the temperature is increased from \SI{0.05}~{K} to \SI{0.9}~{K} (from left to right---the red line indicates data obtained at the fridge base temperature of \SI{0.05}{K}, and the black lines are for $T$ from \SI{0.1}{K} to \SI{0.9}{K} in steps of \SI{0.1}{K}). Traces are offset to the right for clarity.}
\label{tempdep}
\end{figure}

Fig.~\ref{tempdep} shows the temperature dependence of the conductance $G$ as a function of split-gate voltage $V_{\rm SG}$ for device B.
The step-like features in conductance are thermally smeared with increasing temperature.
We first address the non-exact conductance quantisation at low $T$. Data are corrected for a constant series resistance ($R_{\rm S}=\SI{110}{\Omega}$) measured at zero gate voltage. This does not lead to a constant vertical spacing of $2e^2/h$ between plateaus, instead the spacing is less than $2e^2/h$ near pinch-off and increases to close to $2e^2/h$ as $G$ increases [also seen in Fig.~1(b) of main article]. This indicates that the series resistance increases as the conductance decreases. The correction overestimates the voltage drop across the device, and underestimates the device conductance. Therefore, this simple correction does not manage to align all the quantised plateaus with exact integer multiples of $e^2/h$. The region near the wire will be affected by the gate voltage and its series resistance will increase as $V_{\rm SG}$ goes more negative, reducing the electron density and hence the elastic mean free path $l_{\rm e}$ because of reduced screening. $l_{\rm e}$ is lower than in GaAs because of the higher effective mass and may become short enough for collisions with impurities to occur within the region affected by the gate, making the series resistance change more than it does for electrons in GaAs.
It is therefore likely that the apparent suppression of the $4e^2/h$ plateau is the result of an incorrect series resistance, and the plateau should be closer to $4e^2/h$. Moreover, $l_{\rm e}$ decreases with increasing temperature, and this is likely to further increase the series resistance near the channel, as seen in Fig.~\ref{tempdep}. These combined factors make temperature-dependent studies harder than in GaAs. Ignoring the uncertainties in series-resistance calibration, the second plateau ($N=2$) degrades much faster than the `1.7' structure, which is consistent with the behaviour of the 0.7 structure in GaAs, as shown by the arrows in Fig.~\ref{tempdep}.

\subsection{Measurement of electron effective mass}
\begin{figure}[h]
\includegraphics[width=0.7\columnwidth]{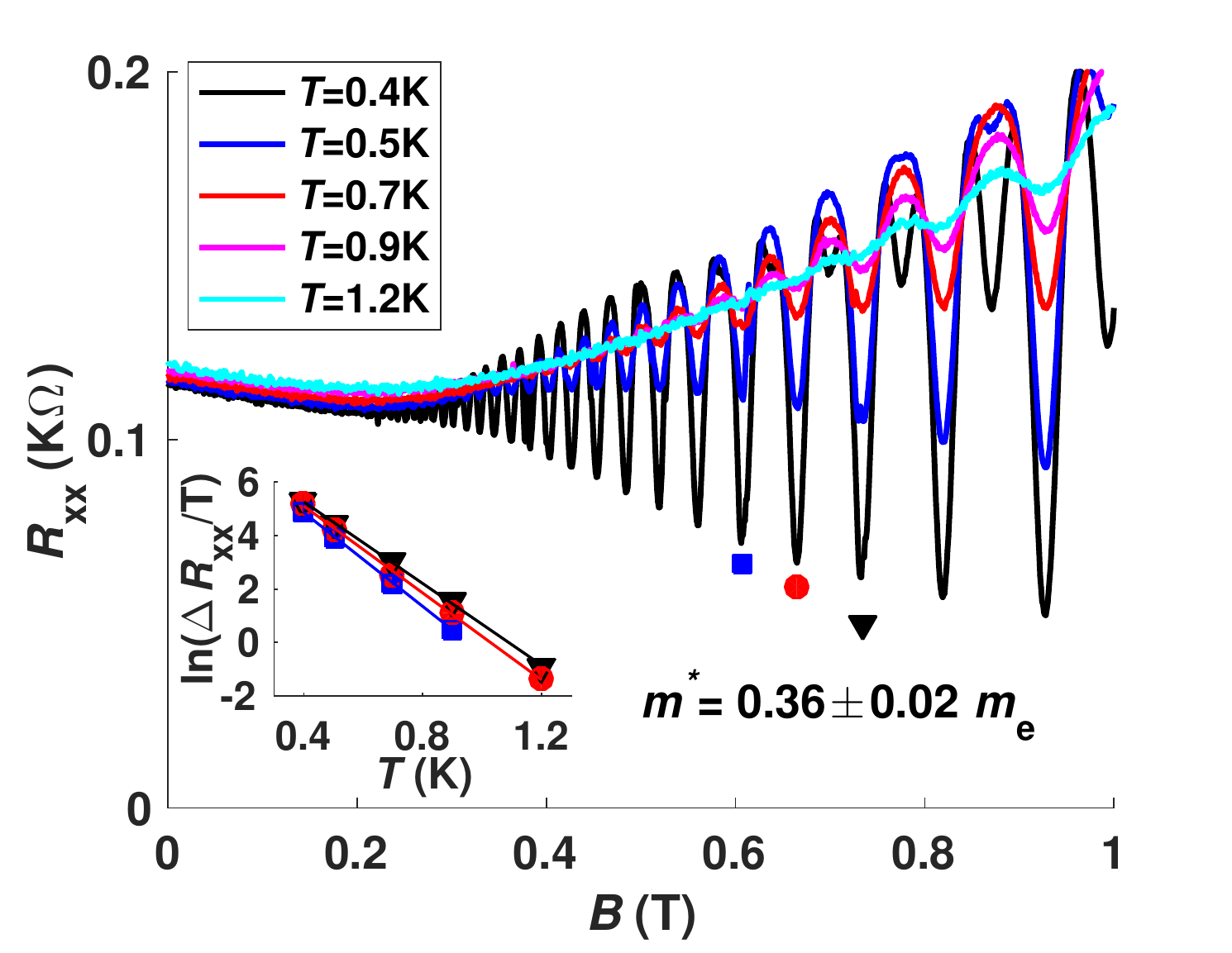}
\caption{Temperature dependence of Shubnikov--de-Haas oscillations from $T=0.4$ to \SI{1.2}{K}. The inset shows the fitting of ln$(\Delta R_{\rm xx}/T)$ as a function of $T$ at fixed magnetic fields, and $m^*$ is estimated to be 0.36$\pm$0.02\,$m_{\rm e}$.}
\end{figure}
We measure Shubnikov--de-Haas oscillations in the 2DEG resistance $R_{\rm xx}$ in the Hall bar containing devices A and B, with all gates grounded. Then the amplitude of Shubnikov--de-Haas oscillation $\Delta R_{\rm xx}$ it fitted to the Dingle expression \cite{coleridge1991dingle}
\begin{equation}
\Delta R_{\rm xx}\propto \frac{\chi}{\sinh\chi }e^{-\pi/(\omega_{\rm c}\tau_{\rm q})},
\end{equation}
where $\chi=2\pi^2k_{\rm B}T/(\hbar\omega_{\rm c})$, $\omega_{\rm c}=e B /m^*$, $\tau_{\rm q}$ is the quantum scattering lifetime (which we assume is temperature-independent over this measurement range) and $T$ is the temperature. The effective mass $m^*$ obtained from this fit is around 0.36\,$m_{\rm e}$ in this MgZnO/ZnO heterostructure, higher than it is in bulk ZnO (around 0.3\,$m_{\rm e}$), where $m_{\rm e}$ is the bare electron mass. In device C, which is fabricated from a MgZnO/ZnO heterostructure with high electron density, $m^*$ at $V_{\rm SG} = 0$ should be close to the bulk value.
\subsection{Asymmetric bias measurements}
\begin{figure}[h]
\includegraphics[width=0.7\columnwidth]{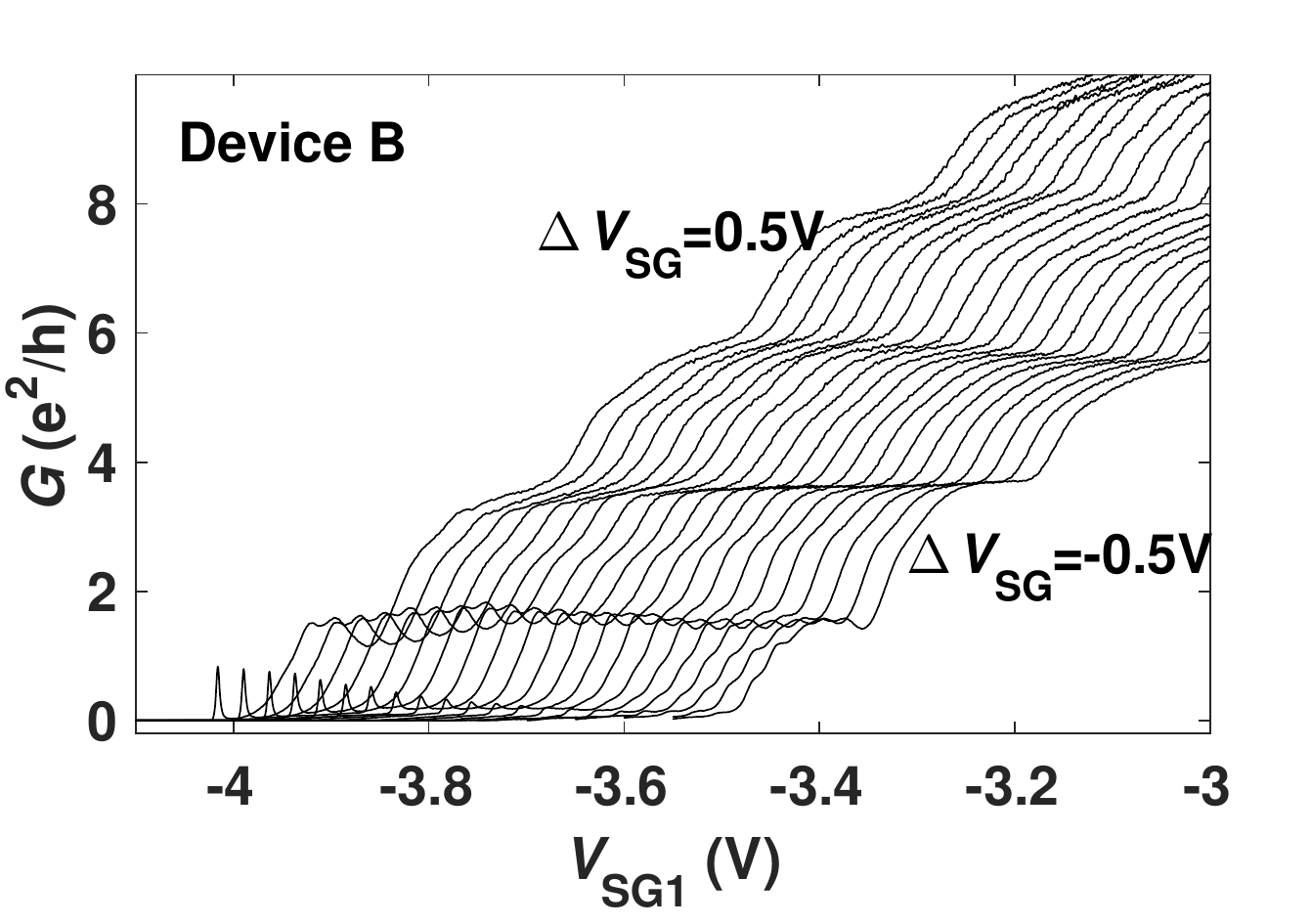}
\caption{Conductance $G$ as function of $V_{\rm SG1}$ as an asymmetric bias is applied to the gates. The difference in voltage on the two split gates is defined by $\Delta V_{\rm SG}=V_{\rm SG1}-V_{\rm SG2}$ changes from \SI{0.5}{V} to \SI{-0.5}{V} in steps of \SI{50}{mV} }
\label{asymmetry}
\end{figure}
Fig.~ \ref{asymmetry} plots the conductance of the 1D wire in Device B with asymmetric bias on quantum point contacts. The quantised plateaus and $N.7$ structures stay reasonably constant as the wire is shifted laterally by an asymmetric bias on the QPC gates. In contrast, the resonant peaks due to CB below the first plateau vary significantly. CB-type features are not independent of position since the disorder potential changes as the wire moves relative to impurities. However, the wire position should not significantly affect either 1D quantisation or interaction effects inherent in the 1D system such as the 0.7 structure (although small variations might occur if the asymmetry alters the wave-function localisation and therefore affects the strength of interactions and 0.7 structure \cite{bauer2013microscopic}). 

\subsection{Other details}

Note that straight lines in intensity plots in the main paper, such as the straight line at $V_{\rm SG} = \SI{-2.9}{V}$ in Fig.~1(c)
(device A) or near $V_{\rm SG} = \SI{-2.6}{V}$ (device B) must come from a random impurity or dot, perhaps under a gate, as they are only very slightly dependent on $B$ or $V_{\rm DC}$ and so provide a conduction path in parallel to the 1D wire.

All DC-bias data shown in the main paper are corrected for series resistance using 
$V_{\rm DC}= V_{\rm DC}^{\rm appl} (1-R_{\rm S} \int_{0}^{V_{\rm DC}}G {\rm d}V_{\rm DC})$, where $R_{\rm S}$ is the series resistance measured at zero gate voltage, and $V_{\rm SD}^{\rm appl}$ is the DC bias applied in the measurement \cite{chen2008bias}.

This MgZnO/ZnO heterostructure also shows dilute ferromagnetic properties with an anomalous Hall effect brought about by spin-dependent electron scattering off localised magnetic moments, which are likely to arise from point defects in epitaxial ZnO with localised unpaired electrons\cite{maryenko2017znoAHE}. This may increase the strength of 0.7-like structure\cite{aryanpour2009ferromagnetic} because this generally appears to be strongly related to spin\cite{thomas1996possible,bauer2013microscopic}. However, the dilute ferromagnetic moments cannot be strong enough to produce full spontaneous electron spin polarisation in the 1D wire, as only plateaus at even multiples of $e^2/h$ are observed at $B=0$. Instead, the dilute ferromagnetism may possibly help to enhance the local spin susceptibility in the channel at $B=0$,\cite{bauer2013microscopic} making it easier for interactions to give rise to the shoulders on each plateau (the `$N$.7' structure) seen, for example, as a pair of vertical lines in Fig.~2(b) in the main paper.


\bibliography{ZnO}
\clearpage

\end{document}